\documentclass[prl,floatfix,twocolumn,showkeys]{revtex4-1}

\usepackage{color} 
\usepackage{graphicx} 
\usepackage{amsmath,amssymb,enumerate} 

\newcommand{\be}{\begin{equation}}
\newcommand{\ee}{\end{equation}}

\begin{document}

\author{Stefano Mostarda}
\author{Federico Levi}
\author{Diego Prada-Gracia}
\author{Florian Mintert}
\email{florian.mintert@frias.uni-freiburg.de}
\author{Francesco Rao}
\email{francesco.rao@frias.uni-freiburg.de}

\affiliation{Freiburg Institute for Advanced Studies, School of Soft
Matter Research, Albert-Ludwigs Universitaet Freiburg, Albertstrasse 19, Freiburg im Breisgau, 79104, Germany.}

\title {Optimal, robust geometries for coherent excitation transport} 

\begin{abstract} 

Coherent transport promises to be the basis for an emerging new technology.
Notwithstanding, a mechanistic understanding of the fundamental principles behind optimal scattering media is still missing.
Here, complex network analysis is applied for the characterization of geometries that result in optimal coherent transport.
The approach is tailored towards the elucidation of the subtle relationship between transport and geometry.
Investigating systems with a different number of elementary units allows us to identify classes of structures which are common to all system sizes and which possess distinct robustness features.
In particular, we find that small groups of two or three sites closely packed together that do not carry excitation at any time are fundamental to realize efficient and robust excitation transport. 
Features identified in small systems recur also in larger systems, what suggests that such strategy can efficiently be used to construct close-to-optimal transport properties irrespective of the system size.

\end{abstract}

\date{\today}

\maketitle

\section{Introduction}

Energy and charge transport are of fundamental importance for technological innovation as well as biological processes such as photosynthesis \cite{Scholes2011,Renger2001}.
If the dynamics is coherent, transport can be enhanced due to constructive interference.
This, however, relies on well defined phase relations which get modified easily if the scattering medium is subject to external or internal sources of noise, even for small perturbations. Consequently, interference is destructive in most ``real world" cases so that the efficiency is reduced to the point where transport might be  completely suppressed
\cite{Lattices1956,Kramer1993,Chin2010,Rebentrost2009c}. 

The
relationship between the detailed spatial configuration of the medium and its functional dynamical properties is subtle \cite{Baumann1986,Renger2009}: two structures with similar geometries can possess strongly different transport properties and, vice versa, two structures with comparable transport properties may not share any evident common geometrical feature \cite{Scholak2011}. Clearly, a mechanistic understanding of the relationship between structure and transport efficiency would be necessary to use quantum coherence as a physical mechanism to develop new technological applications as well as understand photosynthesis at a fundamental level \cite{Li2012,Leegwater1996,Chachisvilis1997}.

A recent application of complex network analysis on a set of randomly arranged excitable sites 
provided a systematic framework to characterize the structural properties of efficient transport \cite{Mostarda2013}. 
%
With much of a surprise, results provided strong evidence for the positive role of a structural motif formed by pair sites that are tightly packed together; although never significantly excited, they assure high transport efficiency and robustness against random displacements of the sites.
This partition into excitation carriers and inactive pairs defines a dynamical separation %
that is reflected in the Hamiltonian, which is approximately composed of two weakly coupled blocks. 
While not necessarily emerging from the same geometrical features, such a dynamical arrangement has been located in some natural light harvesting complexes such as FMO \cite{Brixner2005,Adolphs2006a}.




It is then interesting to understand whether such a active/inactive modular arrangement is a truly general principle or if it is rather a peculiarity of systems of {\it e.g.} specific size.
In this contribution, we therefore consider a paradigmatic system with variable number of randomly disposed excitable sites. Structures with outstanding transport properties are scrutinized, their common geometrical features determined through complex network analyses and their dynamical characteristics studied via inverse participation ratio and eigenvalue distributions. Comparison of the results obtained for different system sizes confirms the presence of specific structural classes for efficient transport that can be differentiated by their robustness properties. This outcome reinforces the idea that tightly packed sites which are not actively involved in the excitation transfer play a fundamental role in the transport, as they make the whole system efficient and robust under perturbations.


\section{Methods}

\subsection{
Tight binding model}
We analyze the transport properties of discrete systems, comprised by a set of $N$ excitable sites that are modeled as two-level systems.
The 
interactions are described by the tight-binding Hamiltonian

\begin{equation} H=\sum_{i\neq j}^N \frac{Jr^{3}_{0}}{|\vec r_i-\vec
r_j|^3}\sigma_i^{-}\sigma_j^{+}\ ,
\label{eq:ham}
 \end{equation} 

where $J$ is the dimensionless coupling constant 
and $\sigma_i^{-/+}$ describe the annihilation/creation of an excitation at site $i$. The interaction rate decays
cubically with the inter-site distance in accordance with dipole-dipole interaction. Within this model, a \emph{structure} is defined by the positions of the $N$ sites. The initially excited site (input) and the site where the excitation is sought to arrive
(output) are located at the diagonally opposite corners of a cube of side $r_0$, while the remaining $N-2$ sites are placed randomly within this cube.

The system is initialized with an excitation on the input site;
transport efficiency is defined as the maximal probability to find the excitation at the output site within a short time interval after initialization
\begin{equation} 
\epsilon=\mathrm{max}_{t\in[0,\mathcal{T}]}|\langle in|\mathrm{e}^{iHt}|out\rangle|^2\ .
\label{eq:emaxtau}
\end{equation}
The states $| in \rangle$/ $| out \rangle$ denote the situation where the input/ output site is excited and all other sites are in their ground state.
In order to target exclusively fast transport that necessarily results from constructive interference, we choose $\mathcal{T} = \frac{1}{10} \frac{2\pi \hbar}{J} \frac{r_{in-out}^3}{r_{0}^3}$, {\it i.e.} a time-scale ten times shorter than the interval associated with direct interaction between input and output sites \cite{Scholak2011,Mostarda2013}. For longer times the excitation would oscillate back and forth between input and output because the dynamics is purely coherent. With a sufficiently short time window, however, only a single oscillation is taken into consideration.

\subsection{Inverse Participation Ratio (IPR)}

Under a coherent dynamics induced by a Hamiltonian of the form given in equation (\ref{eq:ham}) the excitation will get delocalized over the sites of the system.
This delocalization can be quantified in terms of the inverse participation ratio (IPR) defined as
%
\begin{equation}
\mathrm{IPR}(t)=\frac{1}{\sum_{i=1}^{N} q_{i}^2(t)}\ ,
\label{eq:ipr}
\end{equation}
where $q_i$ is the probability for site $i$ to be excited.
A value for the IPR which is larger than $K-1$ implies that the excitation is delocalised over at least $K$ sites. The maximum value of the IPR is $N$, which is obtained in the case of even delocalization over the whole $N$ constituents. On the other hand, if the excitation is completely localized  (e.g. at $t=0$ in our case), the IPR adopts its minimal value of 1.


\subsection{
Efficiency Network}
To unravel the structure-dynamics relationship, we apply a set of tools based on complex networks. Originally, these tools had been developed for the characterization of molecular systems \cite{Rao2004,Gfeller2007}. However, since these methods are designed to analyze large ensembles of configurations, they prove very useful for our present purposes as they allow a systematic classification of structures which lead to exceptional transport.

We generate a complex network where structures with $\epsilon>0.9$ represent the nodes and a link is placed between them if two structures are geometrically similar independently on the specific dynamics of the excitation.
The parameter used to estimate structural similarity depends on the relative distances of the excitable sites of two structures under comparison. The sites are indistinguishable, thus all different permutations of the site labels need to be performed \footnote{There are $(N-2)!$ permutations since input- and output-site are distinguished from the other sites.}.
In addition, a rotational symmetry around the in-out axis and an additional mirror symmetry has to be taken into consideration.
The measure $S$ of similarity between configurations A and B is thus defined as
\begin{equation}
S^2=\min\sum^n_{i=1}  \frac{d_i ^2}{n}\ ,
\end{equation}
where $d_i$ is the difference of the coordinates of the $i-$th site in the two configurations,
and the minimization is performed over all permutations, rotations around the in-out axis and the mirror symmetry.
A link is placed in the network only if $S$ lies below a certain threshold value $S^{*}$ which is going to be discussed in detail in the next sections.


\begin{figure}
\centering
\includegraphics[width=0.475\textwidth]{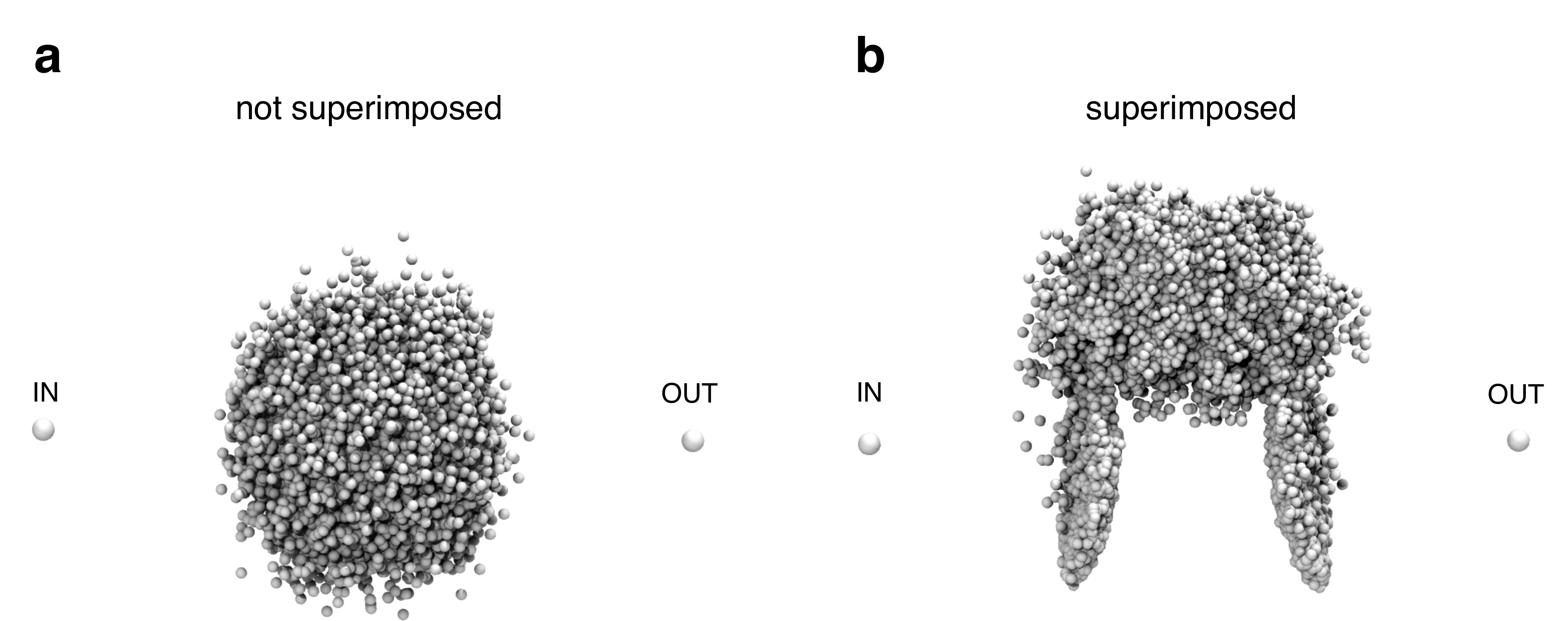}
\caption{The structural superposition algorithm (here on the first cluster with $N=6$, 10731 structures) makes geometrical features emerge from the noise.}
\label{fig:superimposition}
\end{figure}

\subsection{Network Clusterization}
Densely connected regions of the network indicate the presence of groups of structures with common geometrical motifs \cite{Mostarda2013}. We identify these regions using a network clusterization algorithm based on a self-consistency criterion in terms of network random walks, the Markov clusterization algorithm (MCL). The network is in this way split into different clusters comprised of structures with similar sites arrangements \cite{Gfeller2007, Van2008}. The method consists of four steps: 
\begin{enumerate}[(a)]
\item start with the transition matrix $A$ of the network, where each column is normalized to 1; 
\item compute $A^{2}$;
\item take the {\it p}-th power ($p>1$) of every element of $A^{2}$, normalize each column;
\item go back to step (b).
\end{enumerate}
 After some iterations of the MCL, $A$ converges to $A_{MCL}$, where only one entry for each column is non-zero. Clusters are defined by the connected regions of the percolation network. In the limit of $p=1$, only one cluster is detected. On the other hand, the parameter {\it p} is related to the granularity of the clustering process. Large values of {\it p} generate several small clusters.

\subsection{Structural superposition}
A structural representation of the clusters is obtained in the following way: for each cluster, the most connected structure is taken as reference and all the others are superimposed. For each structure, we represent the one obtained with the combination of labeling, rotation and mirror state which minimizes the similarity parameter $S$ (see {\it Network Creation} section). In order to reduce noise, the coordinates of the sites are averaged with the ones from two other structures of the cluster taken at random. Structural rendering is done with VMD \cite{humphrey1996vmd}. An example of the effects of such algorithm is shown in Fig.~\ref{fig:superimposition}.

We follow this procedure to depict all the clusters in Supp. Fig.~\ref{fig:supp:VMDrepr1},\ref{fig:supp:VMDrepr2},\ref{fig:supp:VMDrepr3}.

\begin{figure}
\centering
\includegraphics[width=0.475\textwidth]{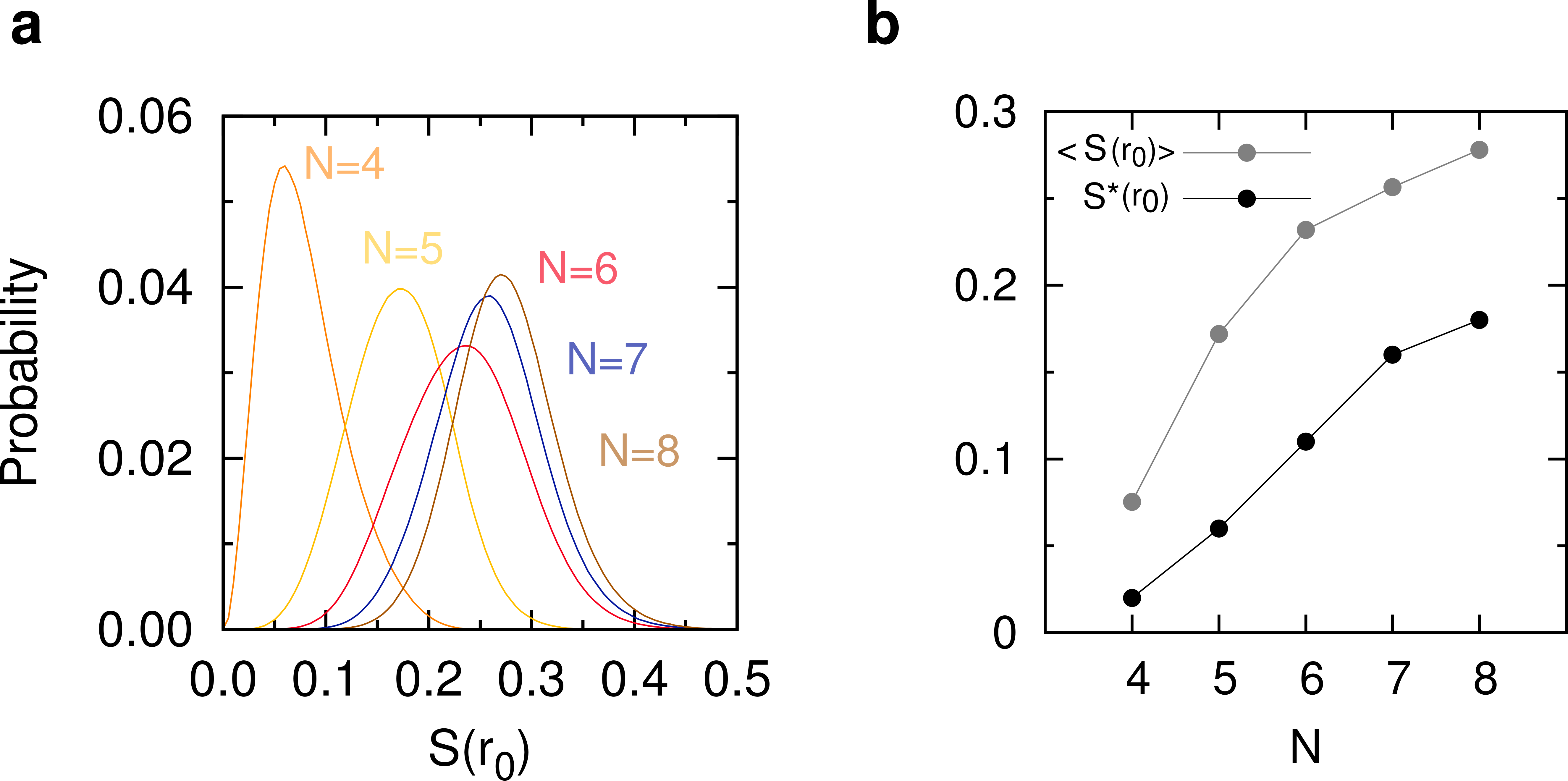}
\caption{Distribution of the similarity parameter $S$ for systems with different number of sites. (a) The average similarity between two given structures decreases ($S$ increases) with the number of sites. (b) The chosen cutoffs (in grey) show a dependence on N that is similar to that of the average similarity $S$ (in black).}
\label{fig:Sdistrib}
\end{figure}

\subsection{Consistency parameter $C$ }
In order to monitor whether the clusterization procedure is consistent while varying the granularity parameter $p$, we introduce here a ``consistency parameter'' $C$. If the clusterization is accurate, an increase in the granularity breaks big clusters into smaller clusters, without mixing them. In fact, every cluster obtained for a given value of $p$ should be fully (or almost fully) included in only one single cluster generated with a smaller value $p -\Delta p$. If this is the case, the clusterization is consistent and the value for $C$ will be maximal. On the other hand, the worst case is a completely random clusterization: the structures of each cluster for a given $p$ are equally distributed between the $n$ clusters generated with $p-\Delta p$. This would correspond to the minimal value of $C$.

For the computation of $C$, we first perform the clusterization analysis from $p=1.1$ to $p=2.2$ in steps of $\Delta p = 0.1$ (from very low to very high values of $p$). For every step $p$, we calculate for each cluster $\mathcal{I}$ the largest portion $C_{\mathcal{I}}$ of its population $\mathcal{P}_{\mathcal{I}}$ included in a single cluster obtained at $p - \Delta p$. $C$ is then calculated as the average of $C_{\mathcal{I}}$ weighted over the relative populations $\mathcal{P}_{\mathcal{I}} / \sum_{\mathcal{I}} \mathcal{P}_{\mathcal{I}}$. In this way, the maximum value of $C$ is always 1, which corresponds to a perfectly consistent clusterization. The minimum of $C$ at a given $p$ is $1/n$, where $n$ is the number of clusters generated at $p -\Delta p$. To make the value of $C$ independent of $n$, we normalize it such that $1/n$ corresponds to 0 and rescale the $[1,1/n]$ segment linearly to $[1,0]$.


\section{Practical advices for the parameters choice}
Network creation and clusterization depend on two parameters: the similarity cutoff $S^{*}$ which sets the accepted degree of similarity between different structures and the granularity parameter $p$ which determines the degree of coarse-graining in the clusterization. It is important to note that finding the correct value of these parameters for structural comparison is an open and unsolved problem in the broader field of complex systems. Apparently, there is no single {\it right} choice, as those parameters probe the system at different resolutions. Best practice suggests a scanning in parameter space in order to asses the robustness of the observations on a particular data set. In this section, we discuss cut-off choices in some detail.

In Fig.~\ref{fig:Sdistrib}-a the distributions of $S$ are shown for the most efficient structures ($\epsilon>0.9$) obtained for $N=4-8$. Interestingly, two behaviors are present. The case $N=4$ is compatible with an almost homogeneous ensemble, where any two structures are very similar to each other ($S<0.15$ for the 94\% of the links). On the other hand, in the systems with $N=6,7,8$ the number of pairs of compatible structures is instead very small, i.e. the ensemble is deeply heterogeneous ($S>0.15$ for the 92\%, 98\% and 99\% of the links for $N=6,7,8$, respectively). The case $N=5$ shows an intermediate behavior.

\begin{figure}
\includegraphics[width=0.475\textwidth]{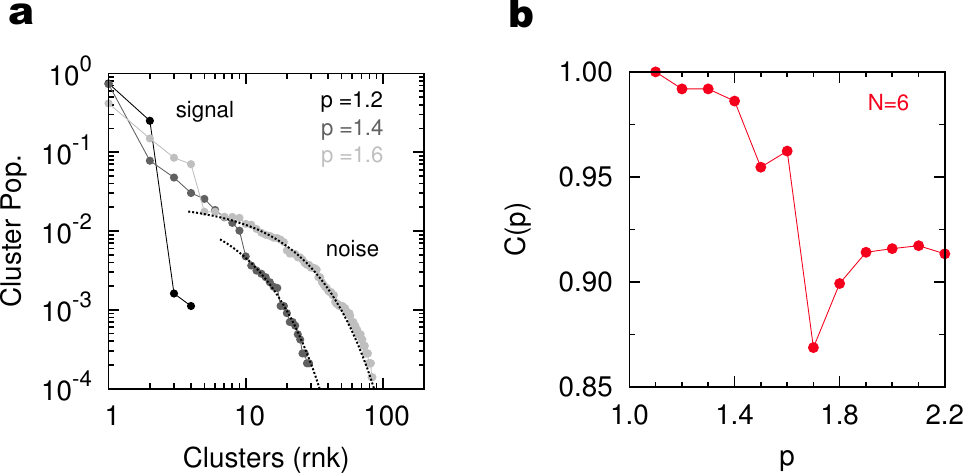}
\caption{Parameters choice in the clusterization procedure for $N=6$. (a) Relative cluster populations for $p=1.2,1.4,1.6$ are shown in black, dark and light grey, respectively. Significant clusters separate from the noise which results in an exponential tail (fitted dashed lines). (b) Consistency parameter $C$ as a function of $p$.}
\label{fig:loyaltyparam}
\end{figure}


If only one system is considered, the cut-off needs just to be self consistent, i.e. the results should not vary too much with $S^{*}$. The problem arises when one wishes to compare different networks, {\it i.e.} different distance distributions. A fixed value for $S^{*}$ for the different cases would create networks with very different connectivities, which makes the comparison very hard. 
In order to set the thresholds in a compatible way, $S^{*}$ is taken as the minimal value of $S$ for which the networks are fully connected (99.9\% of nodes have been considered). The resulting values are $0.02$, $0.06$, $0.11$, $0.16$ and $0.18 \ r_0$ for $N=4-8$, respectively. These values increase in a similar manner as the average value of the distance $S$ (see Fig.~\ref{fig:Sdistrib}-b). $S^{*}$ lies just above the tails of the pairwise distance distributions. Consequently, only the most similar structures are linked together. Lower values of the cut-off would generate a disconnected network, while values too close to the maximum of the distributions would put links between structures that are not very similar. 

For the clusterization process the goal is to separate the bulk of the signal from the statistical noise. To this aim, one can look at the population of the clusters obtained, ranked by decreasing size. Typically, the signal is formed by a small number of populated clusters, while the noise is composed of a large number of small clusters which follow an exponential tail. 

Fig.~\ref{fig:loyaltyparam}-a depicts as an example the results for the $N=6$ case obtained with different $p$. At $p=1.2$ (black curve) the algorithm detects two big clusters with 74.6\% and 25.1\% of the population plus two satellites due to noise with 0.3\% of cumulative population. With $p=1.4$ (dark grey curve) the second cluster at $p=1.2$ splits into eight clusters with smaller relative populations ranging from 1.0\% to 7.8\%, while the noise is composed by the remaining 20 clusters (nicely fitted by an exponential function in Fig.~\ref{fig:loyaltyparam}-a). At $p=1.6$ (light grey curve) only four significant clusters are detected. Their populations are 41.5\%, 14.9\%, 8.5\% and 7.1\% of the total population (cumulatively the 72.0\%), while the remaining 80 clusters have a cumulative relative population of 28.0\% and constitute noise. With even higher values of $p$ the network breaks more and more into small noisy clusters.

These three scenarios show how changing the granularity parameter $p$ leads to different signal to noise 
ratio. This behaviour is not necessarily monotonic: incrementing $p$ at first increases the number of significative clusters up to a maximum after which the noise grows and becomes dominant. However, similarly to the choice of $S^{*}$, our priority is to compare different networks. Therefore, the choice of $p$ which maximizes the signal to noise ratio for each network might not be the best for this purpose. We thus employ the \emph{consistency parameter} $C$, which we calculate for a wide range of $p$ (see Methods for details). This quantity monitors whether the clusterization procedure is accurate and provides a way to consistently compare different networks. 

\begin{figure*}[t!]
  \includegraphics[width=0.99\textwidth]{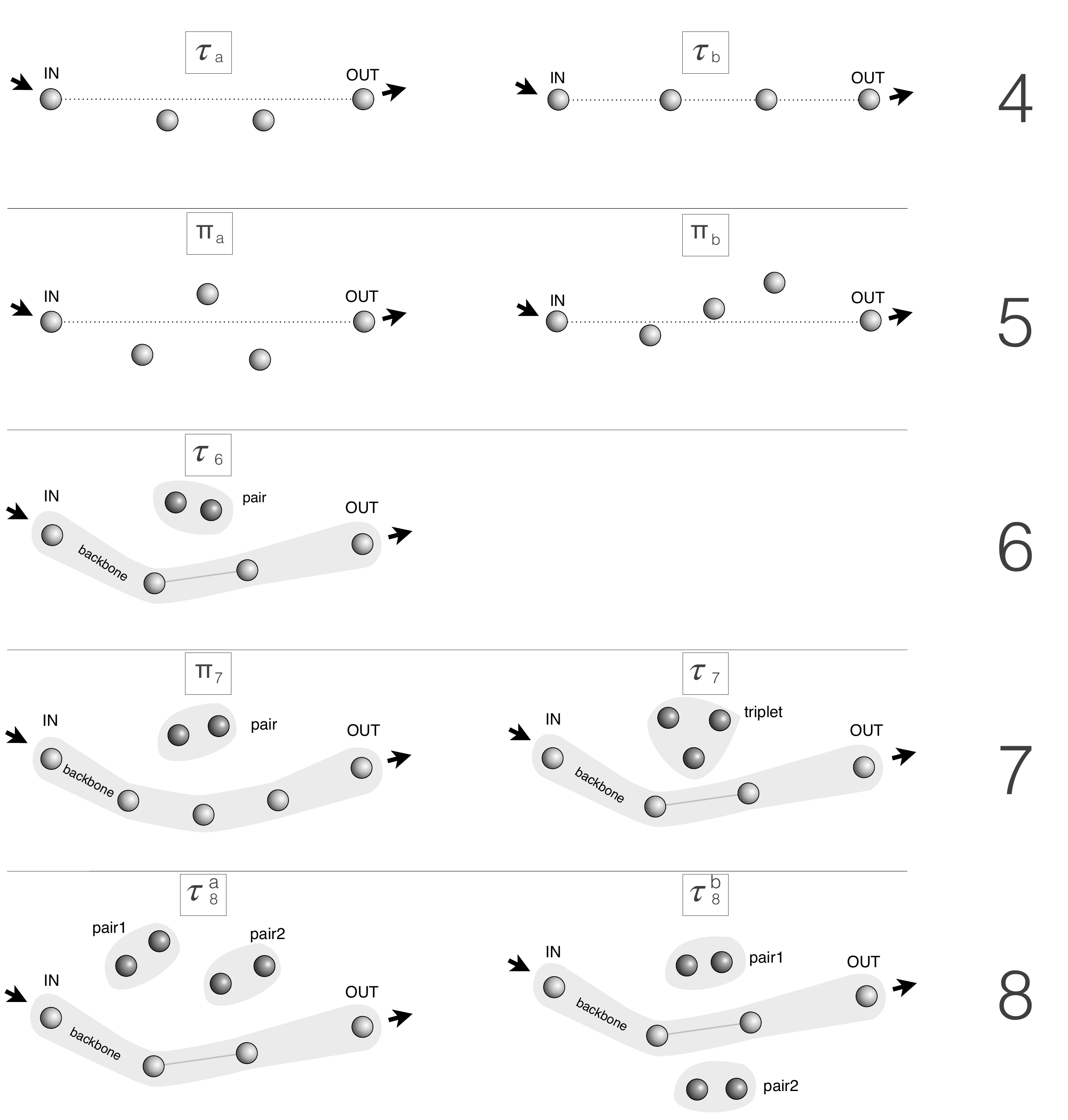} 
  \caption{Schematic representation of the most relevant clusters, organized in rows for systems of increasing $N$. They are the two clusters with $N=4$ ($\tau_a$ and $\tau_b$), the two clusters with $N=5$ ($\pi_a$ and $\pi_b$), the pair cluster of $N=6$ ($\tau_6$), the two clusters with $N=7$ ($\pi_7$ and $\tau_7$) and the two pairs clusters with $N=8$ ($\tau_8^a$ and $\tau_8^b$). For the detailed description of all the clusters, see text.}
  \label{fig:N4cl12}
\end{figure*}

Let us illustrate the behaviour of $C$ for the $N=6$ case in detail as an example (see Fig.~\ref{fig:loyaltyparam}-b): at $p=1$ only one cluster is present, so $C=1$ by definition. At $p=1.1$ we have 2 clusters, but they are obviously both fully contained in the cluster of the former step, thus $C=1$. The first non trivial value of $C$ is at $p=1.2$ where the 4 clusters are quite well identifiable with the 2 clusters at $p=1.1$. There is a slight drop of $C$, but the value $C=0.99$ is sufficiently close to unity to warrant consistency.
This regime is valid up to $p=1.4$, while at $p=1.5$ the value of $C$ drops to $0.95$ (the values of $C$ are normalized); this implies that the clusterization loses some consistency. The biggest drop of $C$ occurs at $p=1.7$, where $C=0.87$ and the consistency of the clusterization process is lost. 

\begin{figure*}[t!]
\centering
\includegraphics[width=80mm]{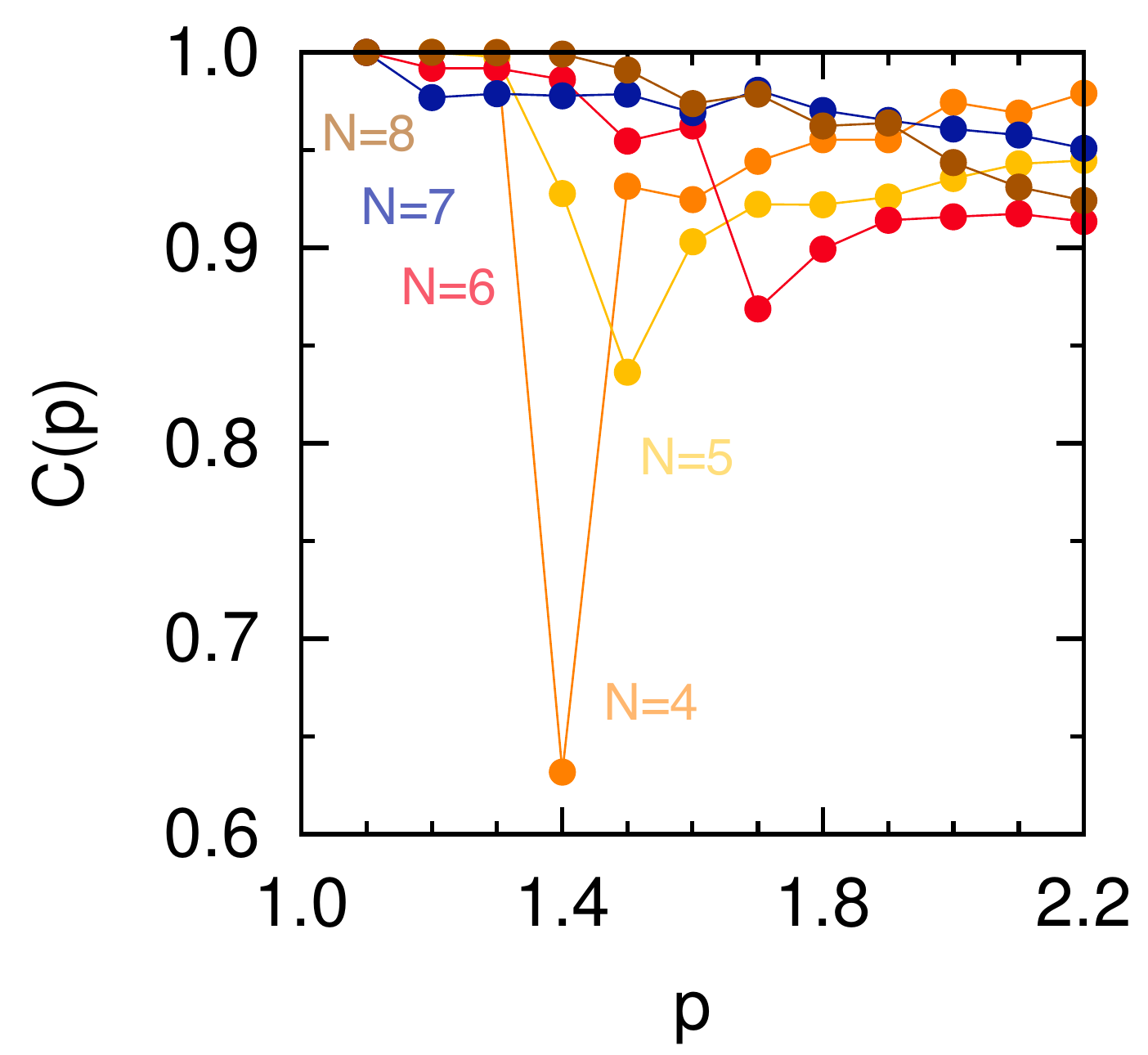}
\caption{Consistency parameter $C$ as a function of $p$ for different values of $N$. The values for $p$ for the clusterization procedure for each $N$ are the highest for which $C(p)=1$.}
\label{figsupp:fidelityparam}
\end{figure*}

For each choice of $N$, $C$ has a different behavior (this can be seen in Fig.~\ref{figsupp:fidelityparam} in Supp.Mat.). This means that we cannot choose a unique value $p$ to use in all the clusterization processes, but we need to investigate case by case the dependence of $C$ on $p$. We then select the highest value of $p$ for which $C=1$ for a given system size $N$: this systematic choice of $p$ allows a first qualitative understanding of the geometrical characterization of the system. The values correspond to $p=1.3,1.3,1.1,1.1$ for $N=4-7$, respectively. For the case $N=8$, the choice of $p=1.4$ obtained following the mentioned criterion leads to a single cluster. This is probably due to the high value of the $S^{*}$ chosen for this system, which creates a more densely connected network, hard to break into clusters ({\it i.e.} a higher value of $p$ would be needed). We therefore increase in this case the value slightly to $p=1.5$.   


\section{Results}

\subsection{Structural characterization of the clusters}
\label{sect:STRCHAR}

\begin{figure*}[t!]
\centering
\includegraphics[width=80mm]{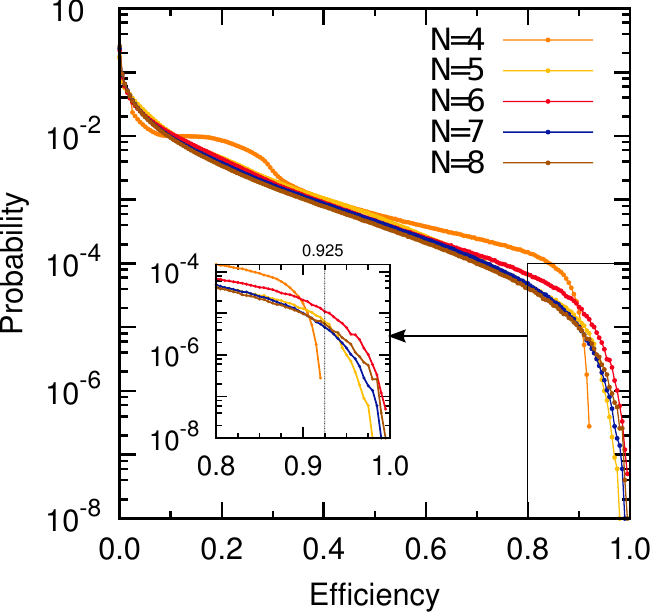}
\caption{Efficiency distribution for systems with different number of sites $N$. As expected from a randomly generated sample, low efficiency are favoured. From $\epsilon = 0.1$ to $\epsilon = 0.9$ the distributions (all but $N=4$) follow an exponential decay, approaching 0 at efficiencies close to 100\%. The inset depicts a zoom at high efficiencies. The efficiency for $N=4$ is never higher than $0.925$ for the chosen time interval (equation (\ref{eq:emaxtau})).}
\label{figsupp:effdistrib}
\end{figure*}

We analyze quantum transport for a large sample ($10^8$) of randomly generated structures with different number of sites ($N=4-8$, see Methods for details). 
The case $N=3$ has not been studied, since it never leads to efficiencies higher than 37\% \cite{scholakphd}. 
Our analysis focuses on structures with $\epsilon>0.9$. 
Within this reduced set, the number of efficient structures is 3530, 7368, 14280, 5896, 6688 for $N=4$ to $N=8$, respectively (in Fig.~\ref{figsupp:effdistrib} in Supp. Mat. the probability of generating efficient structures is shown for $N=4-8$). 

Most of the sets of efficient structures are highly \emph{heterogeneous}, which means that two structures with similar efficiency do not necessarily share any evident common pattern. This structural heterogeneity prevents a straightforward identifications of the geometrical features that are compatible with efficient transport. To uncover these features, we apply the protocol based on complex networks described in Methods. 

\begin{figure*}[t!]
  \centering
  \includegraphics[width=130mm]{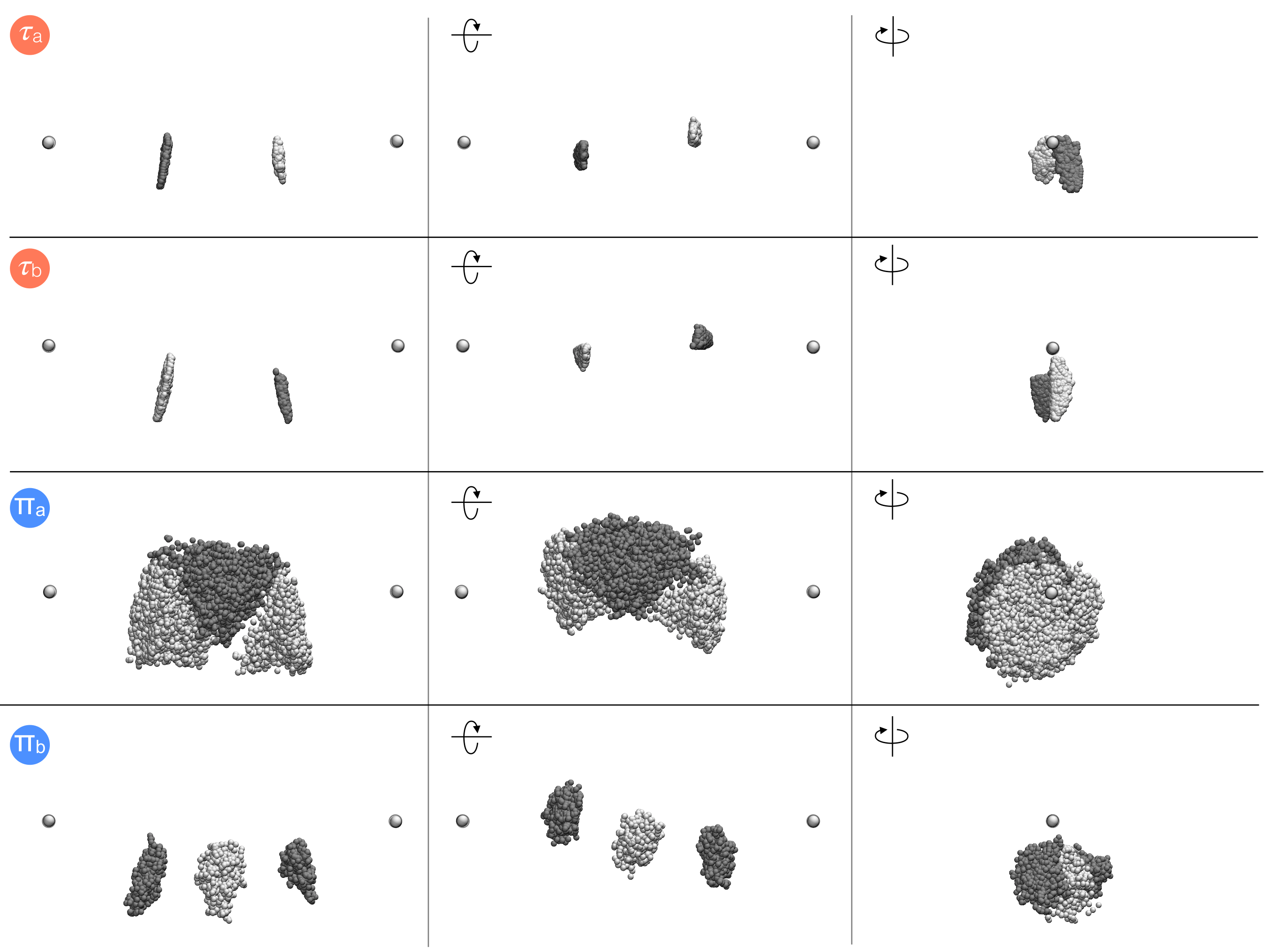} 
  \caption{Superimposition of all structures belonging to $N=4$ and $N=5$. The cluster name is colored according to the class of affiliation (red and blue for $\tau$ and $\pi$, respectively). For clarity, structures are shown in three different orientations.}
  \label{fig:supp:VMDrepr1}
\end{figure*}

\begin{figure*}[t!]
  \centering
  \includegraphics[width=130mm]{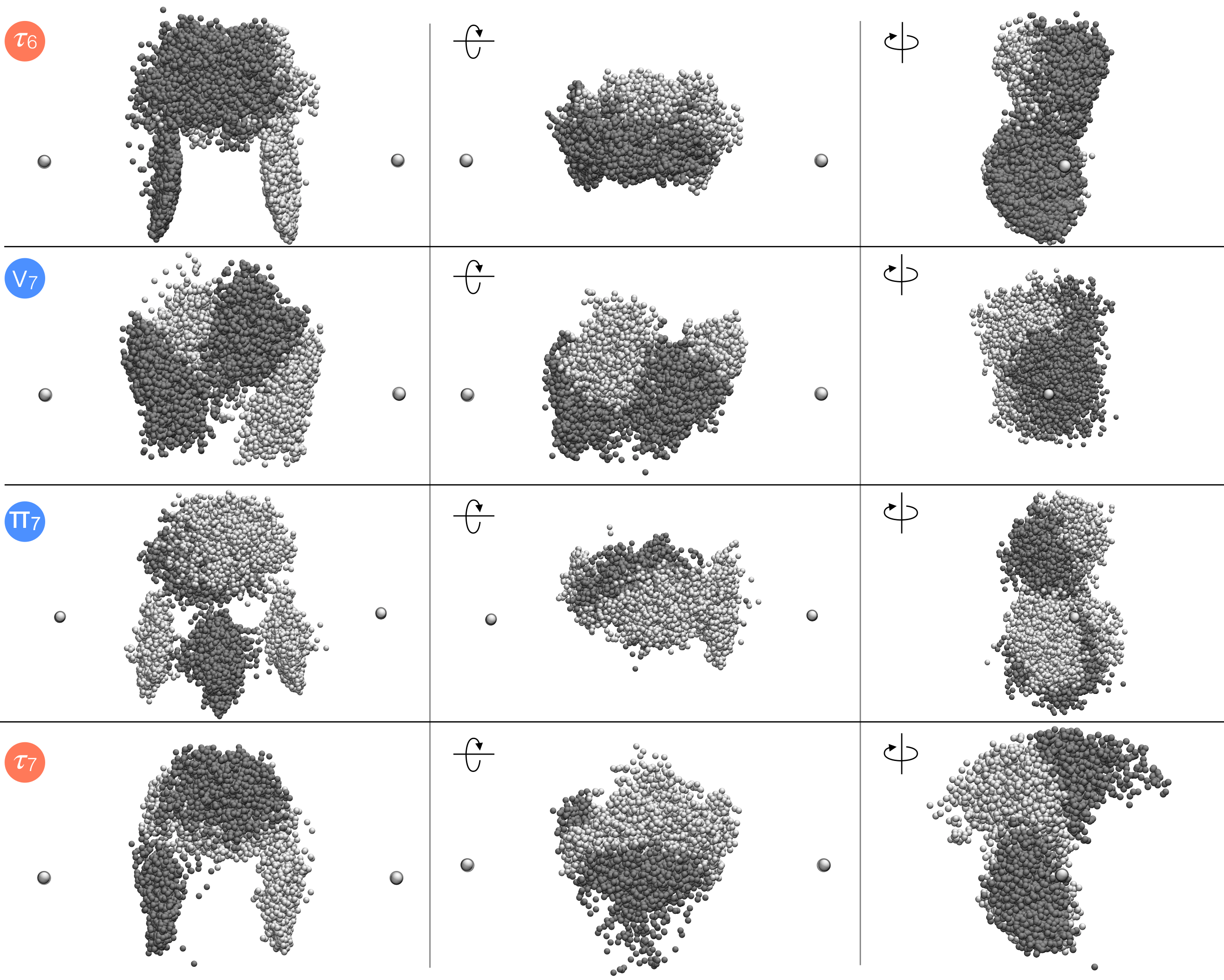} 
  \caption{Superimposition of all structures belonging to $N=6$ and $N=7$. The cluster name is colored according to the class of affiliation (red and blue for $\tau$ and $\pi$ or $v$, respectively). For clarity, structures are shown in three different orientations.}
  \label{fig:supp:VMDrepr2}
\end{figure*}

\begin{figure*}[t!]
  \centering
  \includegraphics[width=130mm]{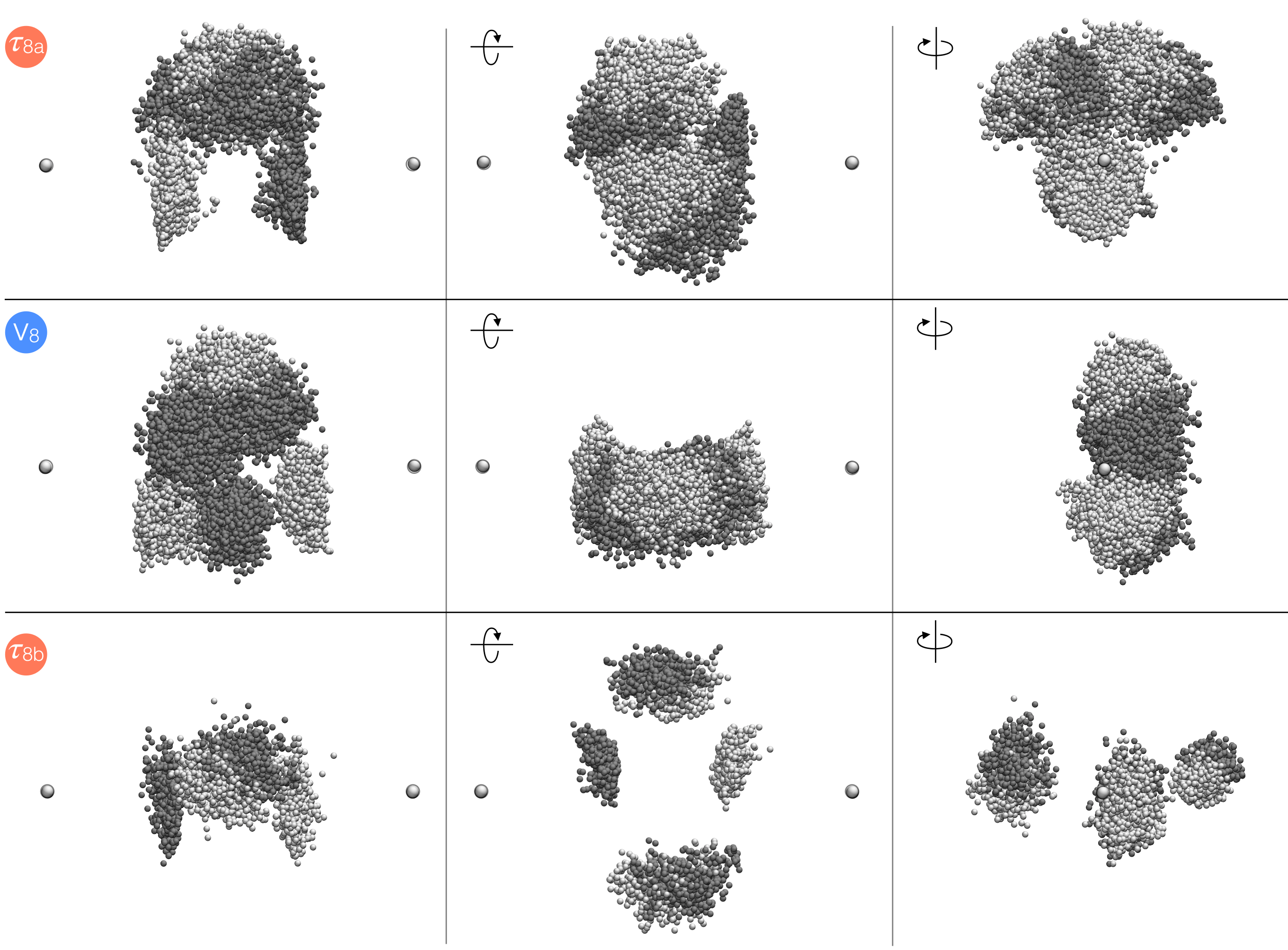} 
  \caption{Superimposition of all structures belonging to $N=8$. The cluster name is colored according to the class of affiliation (red and blue for $\tau$ and $\pi$ or $v$, respectively). For clarity, structures are shown in three different orientations.}
  \label{fig:supp:VMDrepr3}
\end{figure*}

All the clusters we identify are shown in Supp. Fig.~\ref{fig:supp:VMDrepr1},\ref{fig:supp:VMDrepr2},\ref{fig:supp:VMDrepr3}. A sketch of the most relevant ones is depicted in Fig.~\ref{fig:N4cl12}.


According to the network analysis, systems with $N=4$ and $N=5$ are quite homogeneous, and few geometries are compatible with efficient transport. The configurations obtained are shown in Fig.~\ref{fig:N4cl12}. In both cases the clusterization algorithm gives two clusters, where the intermediate sites of the less populated cluster are more strongly aligned than the others. 

For $N=4$ we label the two clusters $\tau_a$ and $\tau_b$ (first row of Fig.~\ref{fig:N4cl12}); they represent 63.1\% and 36.9\% of the total population, respectively. In both cases the four sites are equidistant from each other. However, in the latter case, the two intermediate sites are arranged along the input-output axis while in the former case they are slightly offset.

For $N=5$, the situation is similar, but there is an extra intermediate site. Two clusters, named $\pi_a$ and $\pi_b$ (second row of Fig.~\ref{fig:N4cl12}), represent 83.5\% and 15.8\% of the total population (the remaining 0.7\% is noise). In this case there is a slight deviation from an equidistant distribution of the inter-site distances. In the cluster $\pi_b$ the 3 intermediate sites are aligned along an axis which is rotated with respect to the in-out one; in $\pi_a$ these three sites form a triangle.

\begin{figure*}[t!]
  \includegraphics[width=0.99\textwidth]{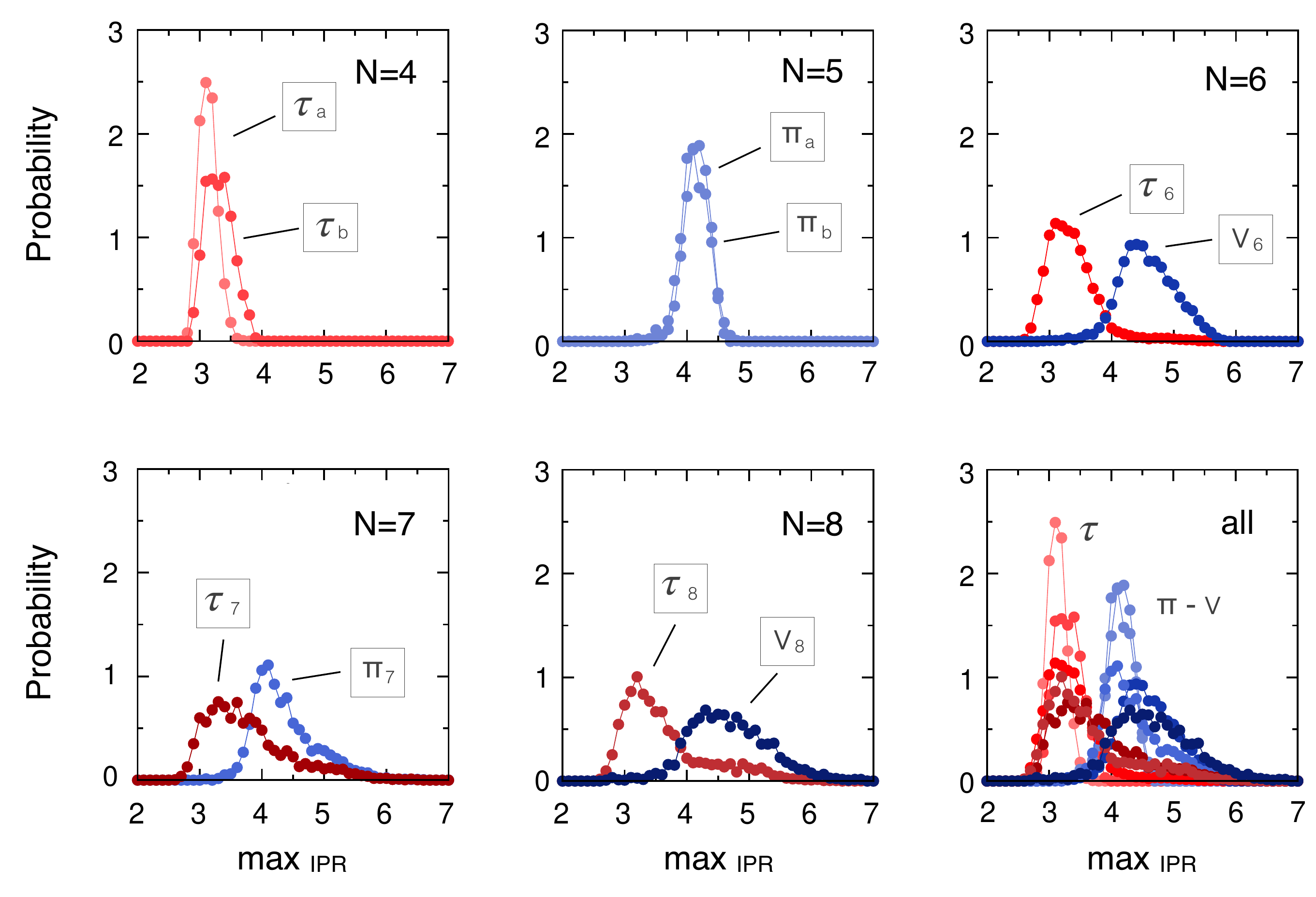} 
  \caption{Distributions of the maxima in $t\in[0,\mathcal{T}]$ of the inverse participation ratios (IPR(t), defined in eq.(\ref{eq:ipr})) for different number of intermediate sites.}
  \label{fig:inverseparticip}
\end{figure*}

Interestingly, the structures found for $N=4$ and $N=5$ constitute the building blocks for the higher dimensional cases (see Supp. Fig.~\ref{fig:supp:VMDrepr1},\ref{fig:supp:VMDrepr2},\ref{fig:supp:VMDrepr3}). In fact, systems with $N>5$ present a higher degree of heterogeneity and a prototypical modular structure. The first module is comprised by four/five sites approximately lined up along the in-out axis; this defines a structural backbone. In all cases this module is compatible with either $\tau_a$/$\tau_b$ or $\pi_a$/$\pi_b$. The second module is essentially formed by the remaining two to four sites, organized in tightly packed \emph{pairs} or \emph{triplets}. Backbone sites are approximately equally spaced between input and output with typical inter-site distances of around $0.50-0.60 \ r_0$, depending on the specific case. Pair/triplet sites instead are always very close to each other with a inter-site distance of around $0.25 \ r_0$, depending on the particular organization. It is worth noting that the backbone arrangement of all mentioned systems is symmetric under input output inversion \cite{Mostarda2013,Walschaers2013}.

The organization of $\tau_b$ constitutes the backbone of the most populated cluster of $N=6$, $\tau_6$ (75.1\%, third row of Fig.~\ref{fig:N4cl12}). In this cluster, we identify a pair whose position is less well defined than the position of the backbone sites. A backbone of four sites that resembles $\tau_b$ is present also in the second cluster for $N=7$, $\tau_7$ (46.9\%, fourth row of Fig.~\ref{fig:N4cl12}, on the right), where the remaining three sites lie close together at comparable reciprocal distances, {\it i.e.} they form a \emph{triplet}. This triplet is located more heterogeneously than the intermediate sites, in a region comparable to the one occupied by the pair module in $\tau_6$. 
Lastly, the most populated cluster for $N=8$, $\tau_8$ (58.4\%), is formed by a backbone of four sites as in $\tau_6$ and $\tau_7$, with the remaining four sites organized into two pairs. At an increase of the granularity parameter to the value $p=1.7$, this cluster breaks into two smaller clusters that differ only in the location of the pairs: the more populated cluster $\tau_8^a$ (47.1\%) has the two pairs on the same side of the backbone, where they form a triangle with the two intermediate sites of the backbone (Fig.~\ref{fig:N4cl12}, fifth row right); the two pairs of the smaller cluster $\tau_8^b$ (10.3\%) are instead diametrically opposed with respect to the backbone axis.

Backbones composed of five sites emerge for $N=7$. The most populated cluster $\pi_7$ (53,1\%, fourth row of Fig.~\ref{fig:N4cl12}, on the left), is in fact formed by five sites organized in a backbone geometry similar to $\pi_a$ with an additional pair similar to the case of $\tau_6$. 

With this choice for the granularity parameters, the remaining clusters are less well defined. 
The second cluster $v_6$ for $N=6$ (24.8\%), is composed of heterogeneous structures which are hard to reconduct to a single structural motif. 
In this case, increasing $p$ to $1.4$ separates  this cluster  into $7$ smaller more homogeneous clusters, with sites either disposed on a line or in a sparse manner. A more detailed discussion of the case $N=6$ can be found in \cite{Mostarda2013}. Also the second cluster $v_8$ for $N=8$ (36.9\% of the population) is poorly identifiable. Cluster $v_8$ is in fact composed by a well defined backbone-like module with five sites and the remaining three sites in a sparse configuration. With the chosen granularity value $p=1.5$, the latter module is not compatible with a triplet. Subgroups at higher $p$, but with a very small population,  present a triplet in a similar manner as in $\tau_7$. 

Altogether these results provide evidence for the presence of a modular arrangement in the geometries of efficient structures.



\subsection{Inverse participation ratio}

So far, we have constructed and clusterized a complex network of efficient structures on purely geometrical grounds. Now we move to investigate the dynamics of these structures, to better understand whether the common geometrical features identified correspond to dynamical similarities.

To quantitatively characterize the dynamical behavior of the identified structures, the inverse participation ratio (IPR, see Methods) is calculated at every instant of time for each structure. In Fig.~\ref{fig:inverseparticip} the distributions of the maxima of the IPR within $t\in(0,\mathcal{T})$ from $N=4$ to $N=8$ are shown, divided into clusters. 

Remarkably, the maxima of the IPR spontaneously group into two well defined distributions. The cases $N=4$ and $N=5$ are basically homogeneous, with negligible differences between the two clusters in both systems. The corresponding values lie around $3.3$ and $4.2$, respectively, which means that the excitation is shared between approximately four or five sites. The two values are prototypical for the IPR distributions of clusters with bigger number of sites. In fact, structures in cluster $\tau_6$ have IPR maxima values similar to those in $\tau_a$ and $\tau_b$, while values for structures in $v_6$ have values close to those in $\pi_a$ and $\pi_b$. The distributions for triplet cluster in $N=7$ and the double pair clusters in $N=8$ (red curves) correspond to those for $\tau_a$ and $\tau_b$, while the pair cluster in $N=7$ and the sparse cluster in $N=8$ (blue curves) have the same IPR distribution as $\pi_a$ and $\pi_b$. 

Strikingly, the distribution of IPR supports that indeed the backbones of the $\tau_X$ and $\pi_X$ clusters (where $X$ stands for any $N$) for $N>5$ correspond to $\tau_a$/$\tau_b$ and $\pi_a$/$\pi_b$ respectively, not only from a geometric point of view, but also dynamically. This is evidenced by the excellent overlap of the distributions for different $N$ (bottom right panel of Fig.~\ref{fig:inverseparticip}). 

\subsection{Inactive sites enhanced transport}


The distributions of the IPR maxima reveal that for $N>5$ only a subset of the sites is substantially excited at the same time. In fact, the sites arranged in a backbone and those forming a pair or triplet possess a different dynamical role; while the former carry the excitation actively, sites closely packed together are never significantly populated by the excitation. Such a behavior emerges systematically for all system sizes, such that the pairs identified before for the case $N=6$ \cite{Mostarda2013} are just one example. In the following sections, we will thus refer to the backbone and to the pairs/triplets as to the {\it active} and {\it inactive} modules of the clusters.

\begin{figure}[t!]
  \includegraphics[width=0.44\textwidth]{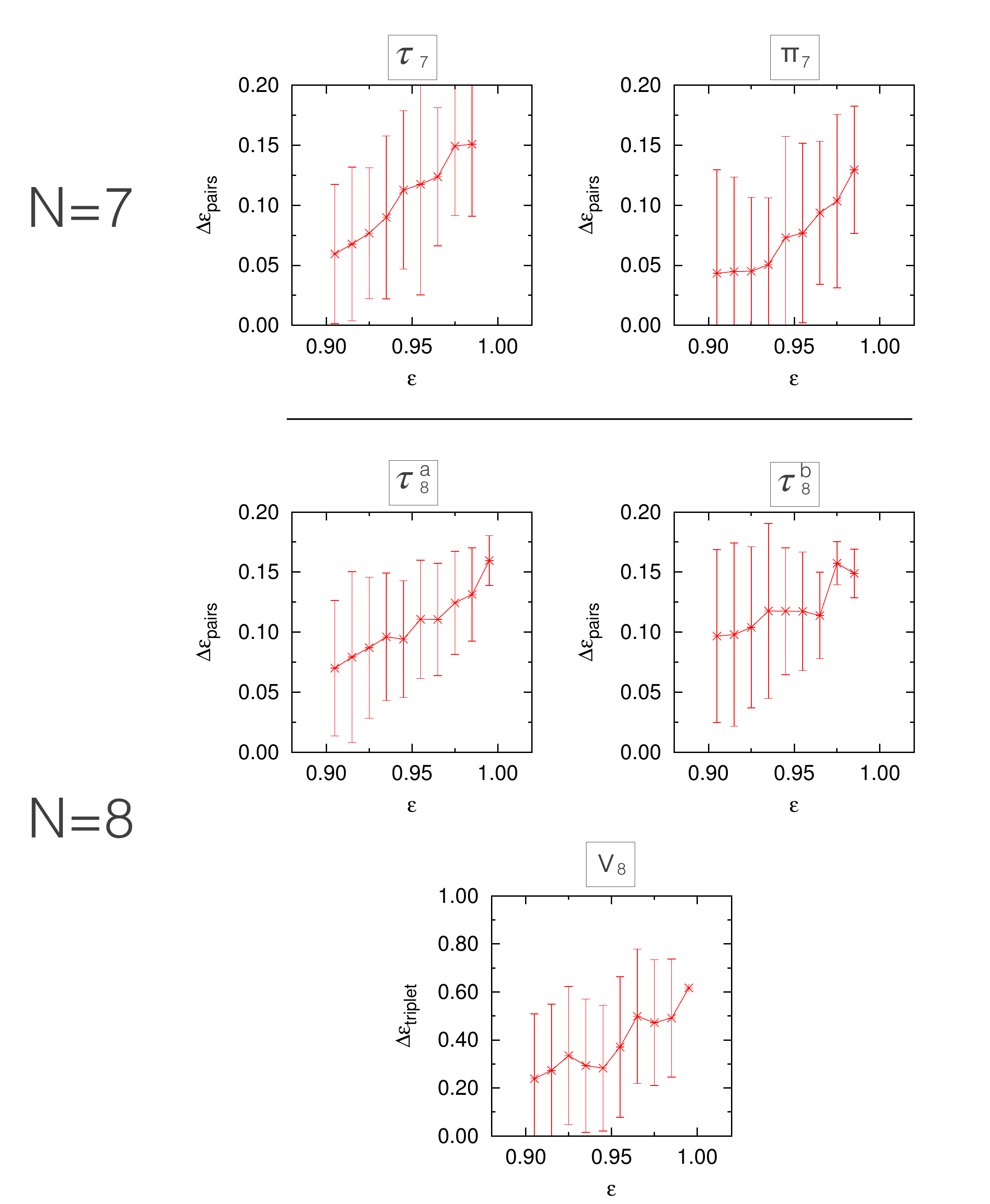} 
  \caption{Efficiency loss upon removal of inactive modules as a function of the original efficiency $\epsilon$ for the clusters with $N=7$ and $N=8$. Error bars are calculated according to the standard deviation. All cases but $v_8$ are compatible with the pair effect found for $\tau_6$.}
  \label{fig:efflossinactivemod}
\end{figure}

Removal of the inactive module results in a systematic efficiency loss. This is shown in Fig.~\ref{fig:efflossinactivemod}, where all clusters, apart from $v_8$, behave similarly: the contribution of the inactive modules is particularly important for the most efficient realizations due to the sensitivity of perfect constructive interference. The loss upon pair removal typically ranges from $0.05$ to $0.15$, depending on the initial value of the efficiency. As we suggested from geometrical considerations, $v_8$ is a very noisy cluster and cannot be considered completely composed of a 5-sites backbone plus a triplet. In fact, removal of these three sites causes an efficiency drop up to 60\%, which indicates that the triplet in $v_8$ cannot be considered an inactive module.

\begin{figure*}[t!]
  \centering
  \includegraphics[width=100mm]{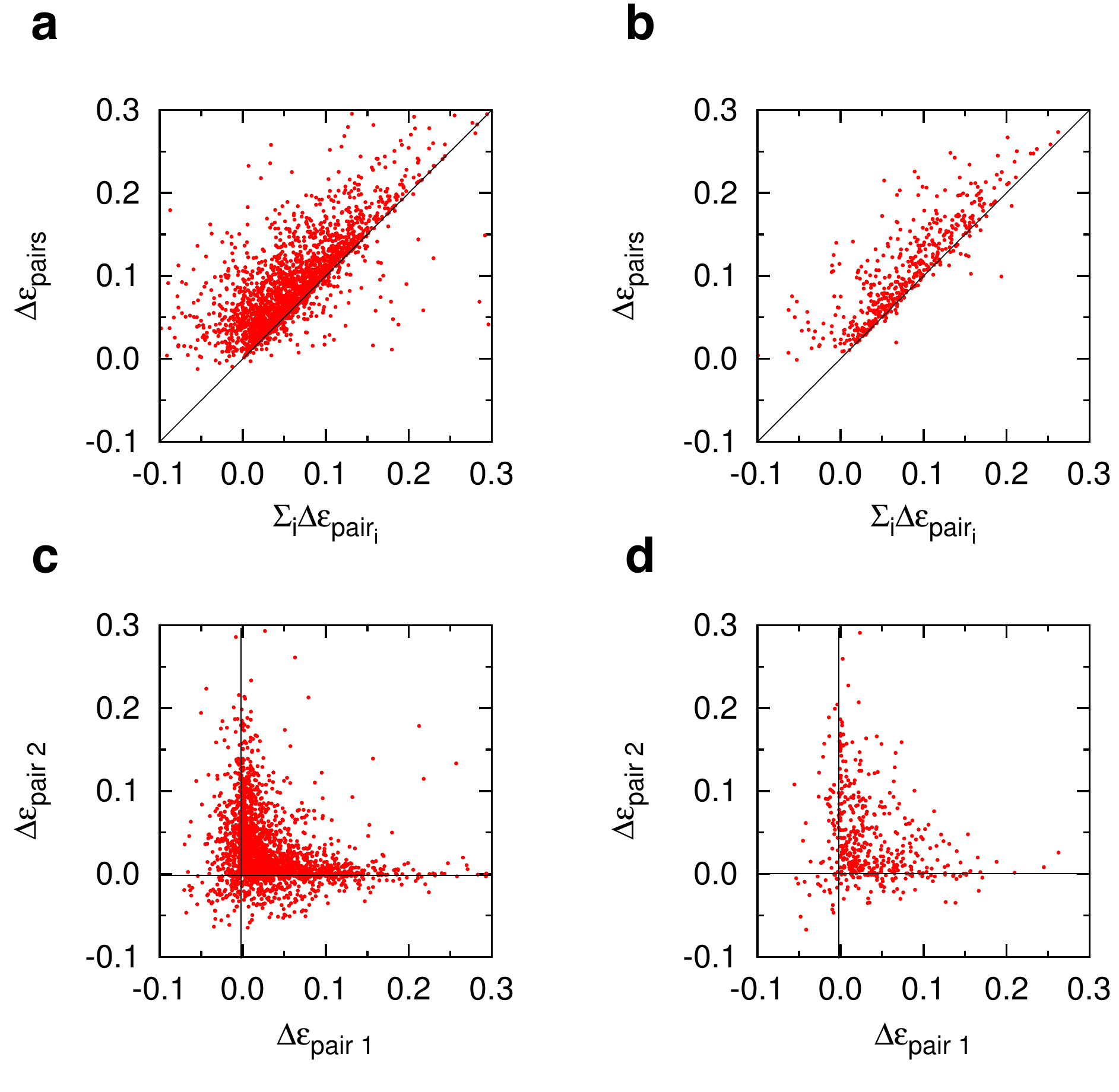} 
  \caption{Comparison of efficiency loss upon removal of pairs for the clusters $\tau_{8}^{a}$ (a,c) and $\tau_{8}^{b}$ (b,d). In order to understand the degree of collectiveness of the pair effect, we compare the sum of the individual losses in efficiency upon removal of each pair with the global loss upon deletion of both pairs at the same time (a,b). These two values are compatible for both clusters. This means that only one pair is responsible for the tuning of the coherence; comparison of the losses upon deletion of individual pairs (c,d) confirms this claim.}
  \label{fig:efflossinactivemodonebyone}
\end{figure*}

While the triplets in the first cluster with $N=7$ play a role similar to the pair in $\tau_6$, it is not obvious whether the presence of two pairs in $\tau_{8}^a$ and $\tau_{8}^b$ is necessary or if only one of them is enough to obtain the same effect. In fact, the two pairs show a small degree of collectiveness, which means that one is dominant and the other one has a close-to-negligible effect (Fig.~\ref{fig:efflossinactivemodonebyone}(c-d) in Supp.Mat.). This is confirmed by the fact that the efficiency loss upon removal of the two pairs at the same time is only slightly larger than the sum of efficiency losses upon removal either pair(Fig.~\ref{fig:efflossinactivemodonebyone}(a-b) in Supp. Mat.).

\subsection{Inactive modules induce eigenvalue shift}

\begin{figure}[t!]
  \includegraphics[width=0.44\textwidth]{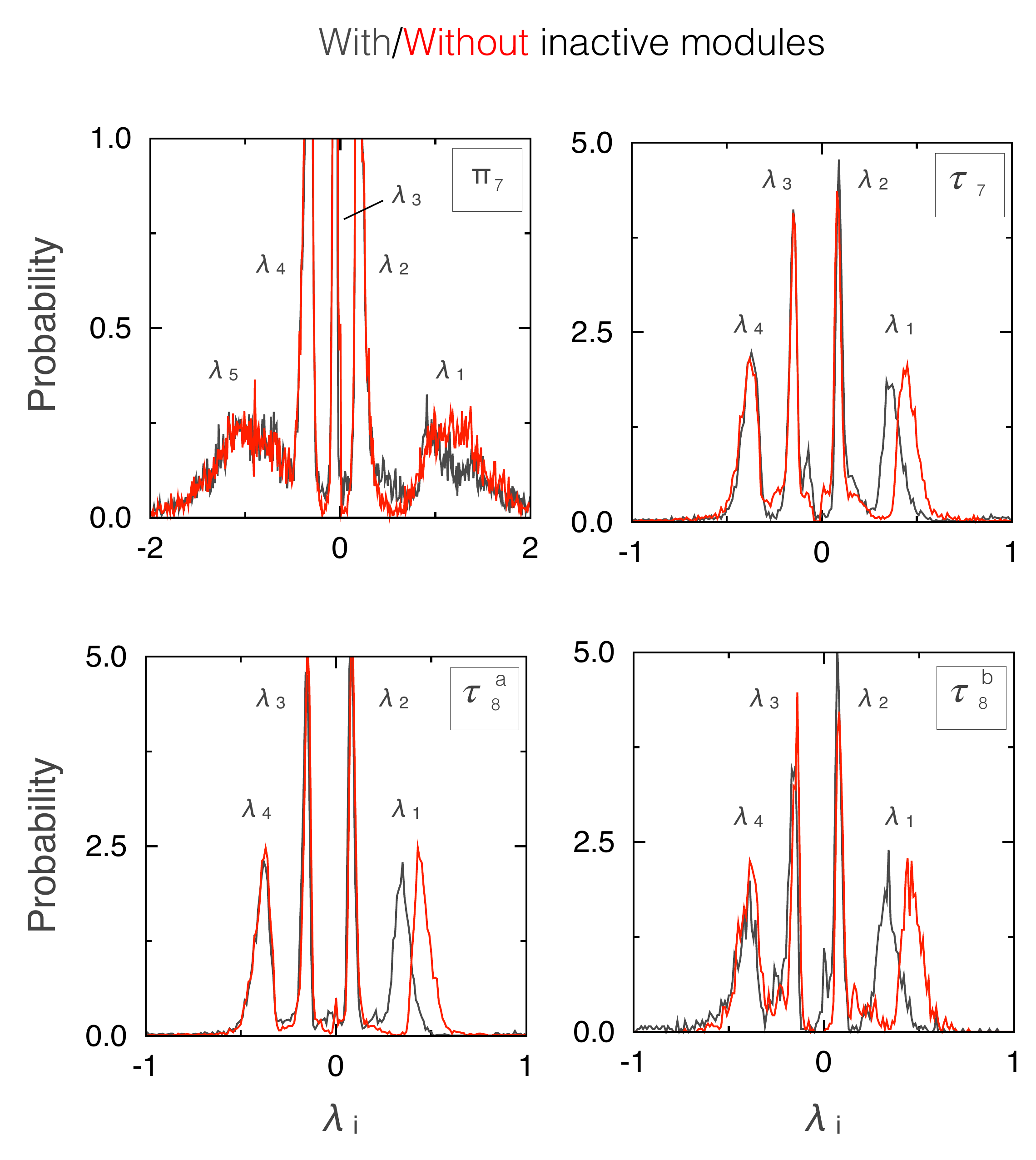} 
  \caption{Shift of the first backbone eigenvalue upon removal of inactive modules. In systems with both $N=7$ (top row) and $N=8$ (bottom row) the removal of the inactive sites causes a shift in only the first eigenvalue. In the former case, the mechanism is similar for both the pair and the triplet.}
  \label{fig:lambdashift}
\end{figure}

The mechanism behind the influence of the inactive modules on the exciton dynamics can be understood from the distribution of the energy eigenvalues with and without the inactive sites as displayed in Fig.~\ref{fig:lambdashift}.
Because only a single excitation is present in the system at any time, there are $N$ energy eigenvalues to study.
Given the weak interaction between the backbone and $k$ pairs or a triplet there are $N-2k$ or $N-3$ eigenstates whose amplitudes are highly localized on the backbone. The amplitudes of the remaining $2k$ or $3$ eigenstates are instead localized on the inactive sites.

The interaction between the backbone and the inactive sites results in a shift of the eigenfrequencies of the former $N-2k$ or $N-3$ eigenstates (denoted by $\mathit{\lambda}_i$ in Fig.~\ref{fig:lambdashift}), such that their differences  are close to integer multiples of a fundamental frequency. With this shift, the excitation is transferred to the output site essentially perfectly after one period of this fundamental frequency. This is true for all the clusters, ranging from the pair and triplet clusters with $N=7$ ($\pi_7$ in Fig.~\ref{fig:lambdashift}) to the clusters with two pairs in the $N=8$ case ($\tau_{8}^a$ and $\tau_{8}^b$ in Fig.~\ref{fig:lambdashift}), what strongly suggests that this mechanism \cite{Mostarda2013} works independently of the system size.

\subsection{Robustness}

The analysis presented so far shows the emergence of two classes of geometrical and dynamical behavior, characterized by an arrangement into active and inactive modules. In the following, we explore this separation with respect to the robustness properties of the various clusters.

\begin{figure*}[t!]
  \includegraphics[width=0.99\textwidth]{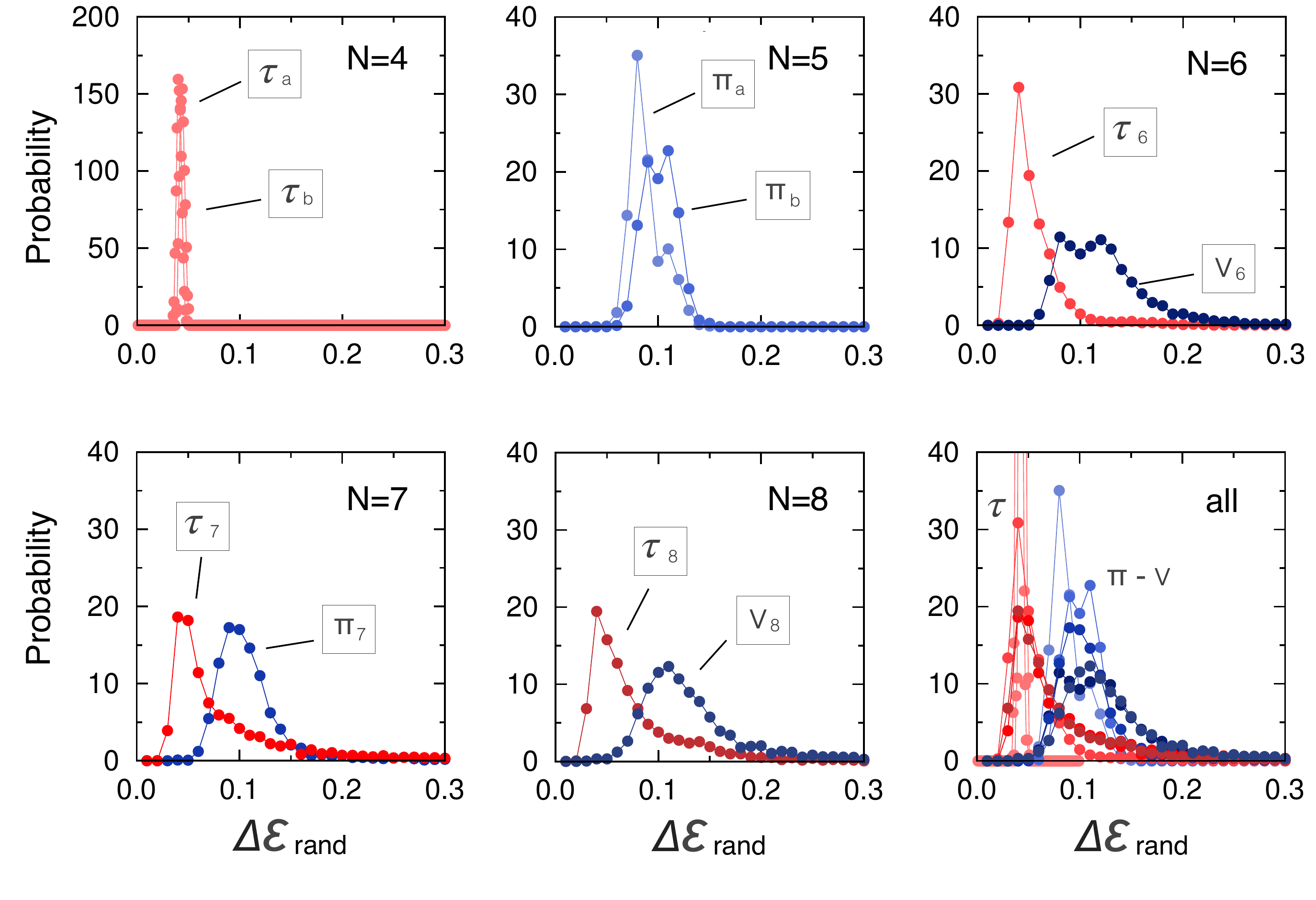} 
  \caption{Distributions of the efficiency loss upon random displacements for different number of intermediate sites. Clusters $\tau_X$ are sistematically more robust than $\pi_X$ and $v_X$. For the case $N=6$, $\tau_6$ presents an average loss of $0.061$, while the same value for $v_6$ is $0.128$. Similarly, $\mathit{\Delta} \epsilon_{\text{rand}}=0.117$ in average for $\tau_7$ and $\mathit{\Delta} \epsilon_{\text{rand}}=0.095$ in average for $\pi_7$. The difference between the robustness of these $N=7$ clusters becomes more evident if one considers that the peak of the second cluster is located at $\mathit{\Delta} \epsilon_{\text{rand}} = 0.045$. For $N=8$, $\tau_{8}$ has $\mathit{\Delta} \epsilon_{\text{rand}}=0.084$ ($\tau_{8}^a$ and $\tau_{8}^b$ have $\mathit{\Delta} \epsilon_{\text{rand}}=0.078$ and $0.092$ respectively), while $v_8$ has $\mathit{\Delta} \epsilon_{\text{rand}}=0.141$.}
  \label{fig:stabilitytot}
\end{figure*}

Transport robustness is probed by random displacements of the individual sites of a structure. With displacements confined to a cube of side $0.05 \  r_0$ centered around the original position of the site, $\mathit{\Delta} \epsilon_{\text{rand}}$ is calculated for each structure as the difference between the original efficiency and the average efficiency obtained from 1000 site-randomizations. In this scheme, structures are kept rigid which corresponds to the assumption that the dynamics occurs on a much faster time scale than low-frequency fluctuations of the entire system (e.g. in the context of biological systems this would be equivalent to large-scale protein breathing).

The distributions of $\mathit{\Delta} \epsilon_{\text{rand}}$ for $N=4-8$ are shown in Fig.~\ref{fig:stabilitytot}. For $N=4$, both $\tau_a$ and $\tau_b$ are very robust under random displacement, the former is slightly more stable than the latter with $\mathit{\Delta} \epsilon_{\text{rand}} = 0.041$ as compared to $\mathit{\Delta} \epsilon_{\text{rand}} = 0.044$. Also in the case of $N=5$, $\pi_a$ and $\pi_b$ present overall a quite similar pattern of robustness: $\mathit{\Delta} \epsilon_{\text{rand}}$ are $0.094$ and $0.107$ for the first and second cluster, respectively. It is however in the clusters obtained for system of size from $N=6$ to $N=8$ that we detect the largest separation in response to random displacements. In all these cases, the loss in efficiency for structures in $\tau_X$ clusters is roughly half of the efficiency of the losses in $\pi_X$ or $v_X$ clusters. Exact values can be found in the caption of Fig.~\ref{fig:stabilitytot}, where 
can be also visually noticed that the two curves separate well from each other in all cases.


Overall, the efficiency loss upon random displacement, which represents the robustness of our randomly generated structures, spontaneously group into two distributions, independently on the number of total sites. This is clearly shown by the overlap of all the curves into a single plot (bottom right panel of Fig.~\ref{fig:stabilitytot}). Two behaviors are present, depending on the number of sites that build the backbone. Clusters whose backbone is composed by four sites (red data in Fig.~\ref{fig:stabilitytot}) show good robustness under random displacement of the sites ($\mathit{\Delta} \epsilon_{\text{rand}}$ peaked around $0.06$), while the efficiency loss of backbones with a larger number of sites (typically $5$) peaks around $0.10$ in all cases (blue data). Poorly defined clusters $v_6$ and $v_8$ share the same response to noise as the $\pi_X$ clusters. Overall, this result suggests that the backbone size is already a good indicator on the robustness of a given efficient structure.

In agreement with the IPR analysis, all the robustness distributions overlap very well (compare the bottom right panels of Fig.~\ref{fig:inverseparticip} for IPR and Fig.~\ref{fig:stabilitytot} for robustness). This provides strong evidence for a clear correlation between robustness, backbone size and inverse participation ratio.


\section{Conclusions}



As shown here, the application of advanced statistical techniques from complex network analysis permit to find a geometrical characterization of efficient structures.
The analysis of efficient transport in systems with a variable number of excitable sites from $N=4$ to $N=8$ highlights the emergence of clear structural signatures related to high efficiency, independently of system size. For growing $N$, a modular arrangement appears. The first is a backbone-like module, typically formed by four or five sites that actively carry the excitation. The remaining sites are arranged in one or more inactive modules composed by tightly packed sites whose function is to tune the eigenvalues of the backbone to realize constructive interference and enhance transport.
This mechanism is statistically dominant: only the 2\% of the structures with $N=7$ or $N=8$ does not possess any inactive module.

Remarkably, common geometrical and dynamical features evidence the recursiveness of these modules. Efficient structures for smaller systems ($N=4,5$) are identified as building blocks for larger structures ($N\geq6$). The addition of inactive modules to these prototypical backbones seem to represent an effective general strategy for the construction of structures in which high efficient transport is achieved by means of constructive interference. 

The analysis presented so far has been performed for a purely coherent case, i.e. without any source of noise. This choice is consistent, because results would not change qualitatively in the presence of incoherent effects. It has been in fact emphazised before that as long as the interest lies in the characterization of {\it fast} transport, which is what motivates the current definition of efficiency equation (\ref{eq:emaxtau}), environmental noise would decrease the efficiency of every structure with no specific distinction \cite{Mostarda2013}, irrespective of the environment considered.

The modularity identified here holds great promise for an explicit exploitation as design principle:
the construction of large optimized system seems feasible if it can be decomposed into smaller, individually optimized units,
whereas a simultaneous optimization over all degrees of freedom easily turns impractical.
Existing aspects of such modularity in actual LHC's underline the feasibility to obtain such optimal structures through evolutionary optimization.
It should thus be expected that the features classified here pave a practical roadmap towards the design of systems that achieve highly efficient transport in a potentially robust fashion.

\bibliography{mostarda2012bisse.bib}

\clearpage
\newpage

{\bf \large SUPPLEMENTARY FIGURES}

\end{document}